Article Type: Technical Article

# An Astro-Animation Class: Optimizing Artistic, Educational and Outreach Outcomes

Laurence Arcadias and Robin Corbet

Laurence Arcadias (animation professor), Maryland Institute College of Art, 1300 W Mount Royal Ave Baltimore, MD 21217, U.S.A. Email: larcadias@mica.edu. ORCID: 0000-0002-5062-0238.

Robin H.D. Corbet (astrophysicist), University of Maryland Baltimore County, 1000 Hilltop Cir, Baltimore, MD 21250, U.S.A.; X-ray Astrophysics Laboratory, Code 662, NASA Goddard Space Flight Center, Greenbelt Rd., Greenbelt, MD 20771, U.S.A.; Maryland Institute College of Art. Email: corbet@umbc.edu. ORCID: 0000-0002-3396-651X.

## Abstract
The authors investigate how teaching art and astronomy together has the potential to inspire new art forms, enhance scientific public outreach, and promote art and science education. The authors teach an astro-animation class at the Maryland Institute College of Art in partnership with NASA scientists. The animations explore science in creative ways. Astrophysicists, educators, students, and the general public were surveyed to evaluate the experiences, and benefits from this project. The responses were very positive - the program is an effective way to stimulate art students to learn science, share an artist's viewpoint beyond the classroom, and engage with the public.

## <1> Teaching Astronomy and Art Together
Stunning imagery helps make astronomy one of the most popular sciences with both the general public and as a subject to study [1]. For example, the NASA "Astronomy Picture of the Day" twitter feed [2] has 1.29 million followers. But astrophysics is also about fundamental physical principles and measurements that are more difficult concepts to grasp which can be an intimidating factor for some people [3]. Incorporating the arts with science in education is a great way to facilitate learning [4--6]. Including artists in the equation opens new possibilities and can generate motivation to explore science from a different angle and connect with a wider audience. ''Science is too important to leave to the scientists, we need voices from the arts and sociocultural disciplines to provoke important debates'' (Stubbs, in [7]).

We are teaching an astro-animation class at the Maryland Institute College of Art (MICA). This undergraduate-level course started in 2014 in the Animation Department and now includes both introductory astrophysics, taught by a scientist, and studio components. NASA-affiliated scientists present their research to the class, on topics such as black holes, supernovae, dark matter, and gravitational waves. Students pick a topic and work in small groups together with a scientist as a mentor. At the end of the semester, the animations are screened at NASA Goddard Space Flight Center's Visitor Center. The animations are freely available at astroanimation.org

and have been broadly used. For further details see [8,9]. The animations have to be accurate to the science but can be whimsical, abstract, poetic, and inspirational rather than directly explanatory (Fig. 1).

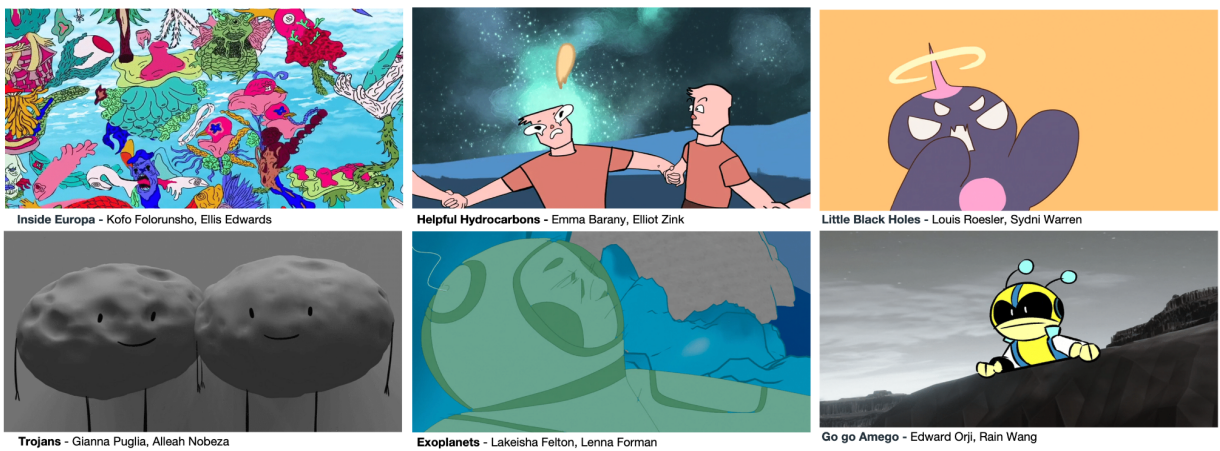

Fig. 1. Examples of the animation styles produced by the students. (compilation © MICA Animation dept.)

Since the class has now been running for several years, we wanted to quantify its effectiveness. We therefore conducted a study examining how animation and astronomy can inspire art making, enhance scientific public outreach, and promote both art and science education. We interviewed astrophysicists, educators, and animation students to document the expectations, experiences, and benefits of our program. We also made a comparison with an art/science class that is offered at the School of the Art Institute of Chicago (SAIC) and conducted interviews with the faculty members and students involved with this class. Based on the results of our research we offer a set of recommendations and examples for other educational institutions wishing to initiate such programs and collaborations.

# <1> Research Program

## <2> Methods - Surveys and Interviews

We administered surveys to scientists, educators, students, and the general public. Surveys were available on paper and online, using scaled answers that measure both the direction and the intensity of opinions. Most people chose to answer the online version. For details see supplemental material.

One survey was administered at the Maryland Science Center (MSC) in Baltimore. We projected our animations on MSC's "Science on a Sphere" [10][11] where many visitors pass by (Fig. 2a). We also screened several of the animations on the same day at the MSC to an audience of middle-school teachers (Fig. 2b). The teachers were not provided with any additional scientific information.

We also carried out surveys at the GSFC Visitor Center and at a public outreach event at MICA associated with an astronomy conference (Fig. 3). The audiences at these events included mentor and non-mentor scientists as well as the general public. Short scientific presentations accompanied both events.

We interviewed MICA students and mentor scientists. The interviews were used to explore more open-ended questions, and to develop a better understanding of the overall responses. We also visited SAIC and conducted interviews with the students and instructors. These interviews were analyzed in the same ways. The interviews were transcribed, and the responses were then coded into a set of categories. This was done both manually, and by using natural language computer processing as described in the supplemental material.

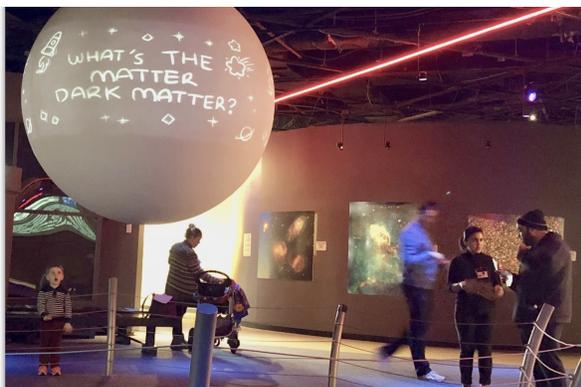 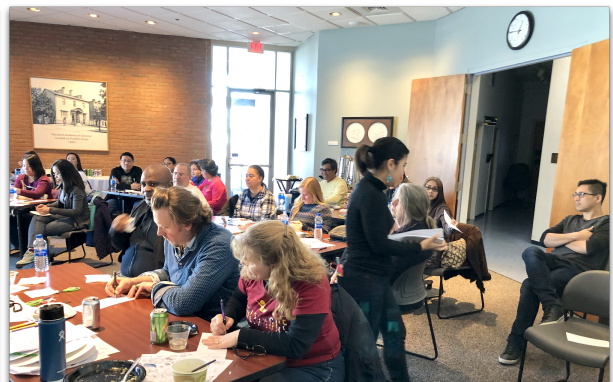

Fig. 2a. Survey with regular visitors at the MSC. (© L. Arcadias)  Fig. 2b. Teachers taking the survey at the MSC. (© L. Arcadias)

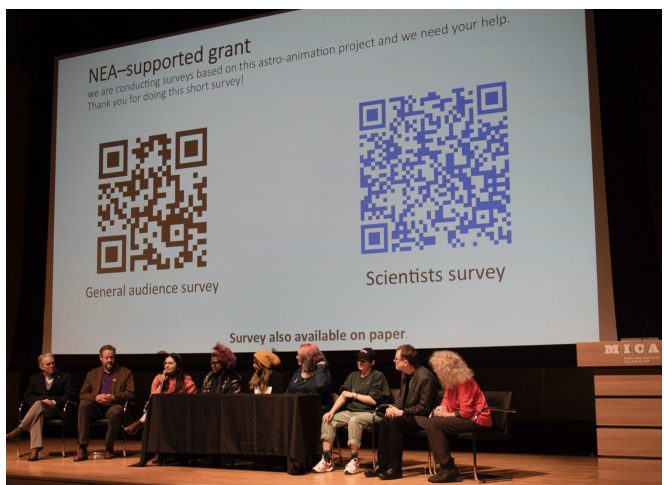

Fig. 3. Surveys at a public scientific outreach event at MICA. (© L. Arcadias)

# <2> ANALYSIS RESULTS

### <3> Highlights of Survey Responses
In Figs. 4 and 5 we show the details of the survey responses. From the scaled answers and the survey comments we identified the following major conclusions

**Fig. 4a. Student's Survey:**

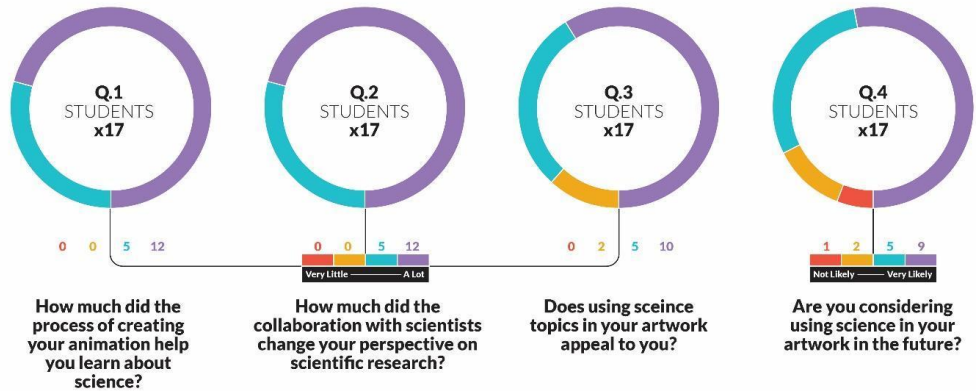

**Fig. 4b. Scientist's Survey**

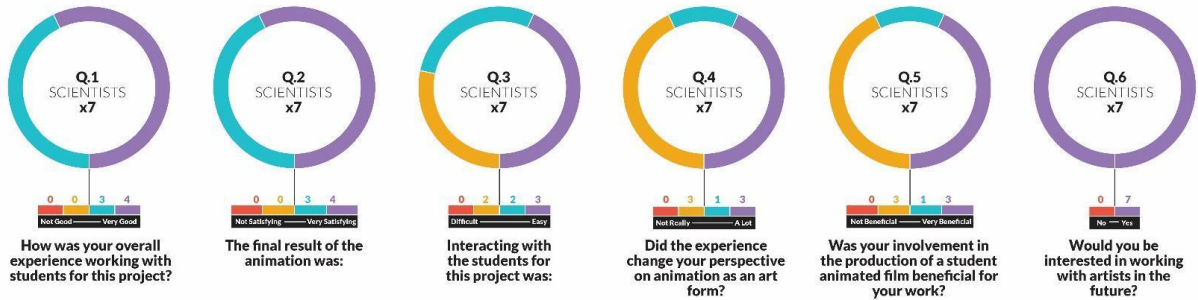

**Fig. 4c. Non-Mentor Survey:**

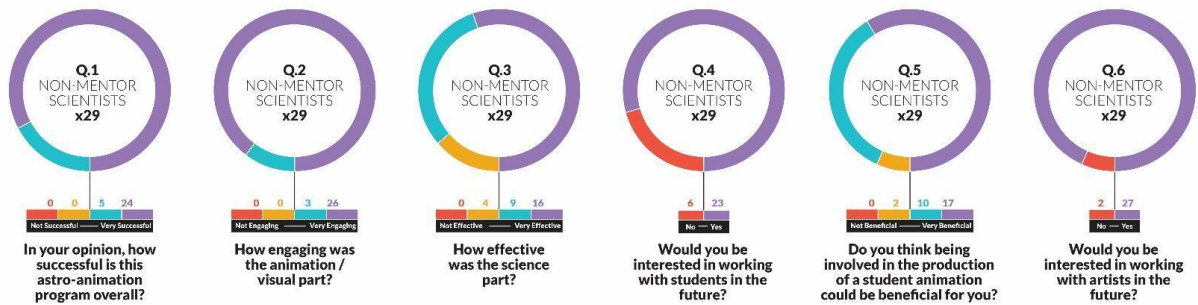

Fig. 4. Survey results and demographic information from surveys of scientists and students. (© L. Arcadias)

## Fig. 5a. Full Audience Survey Results

## Demographic Info

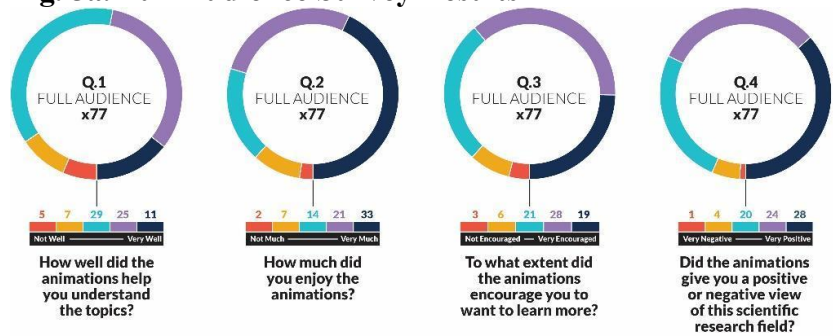
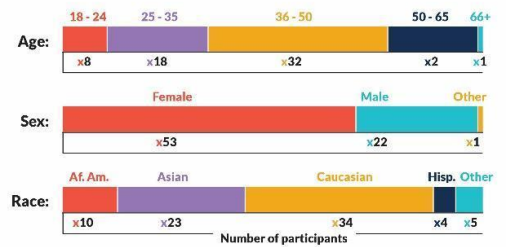

## Fig. 5b. Sub-Groups Survey Results and Demographics:

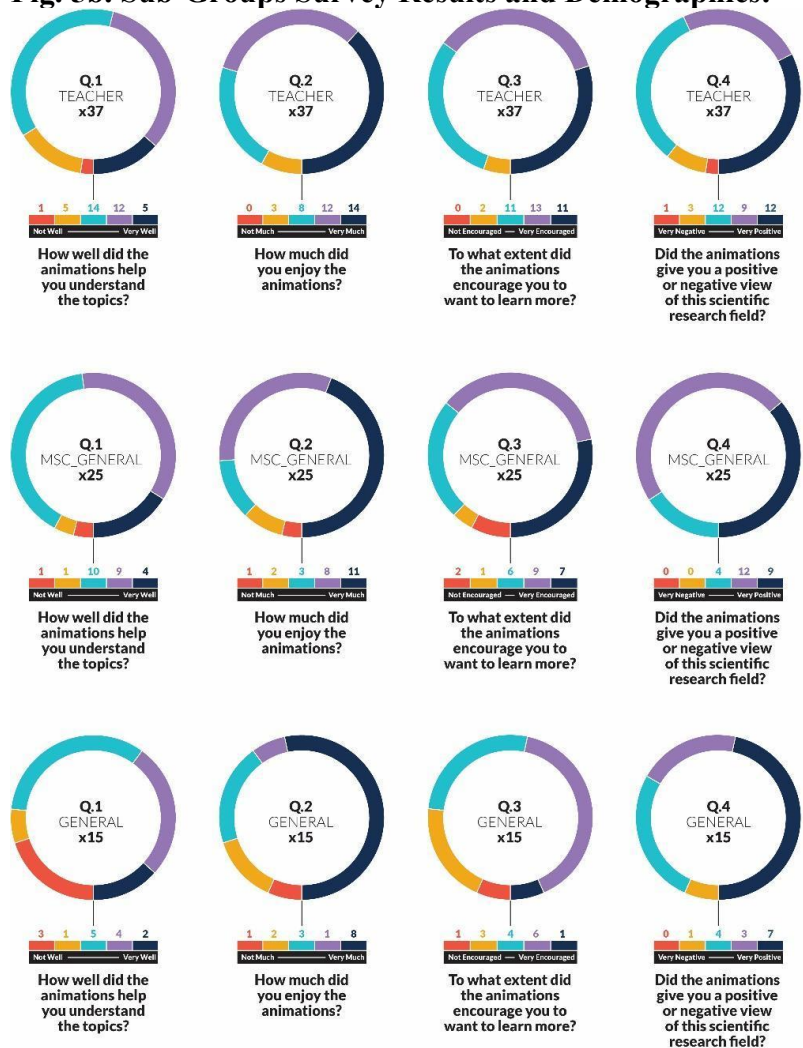
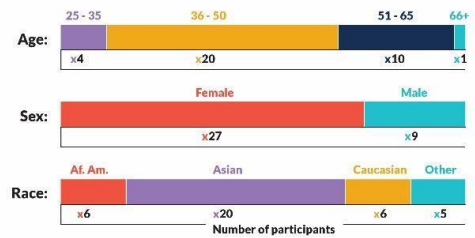
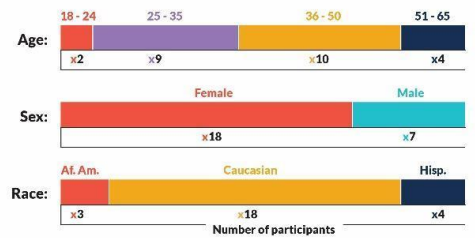
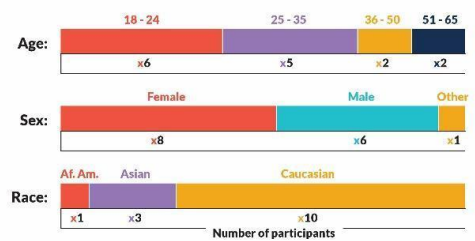

Fig. 5. Survey results and demographic information on survey participants. (© L. Arcadias)

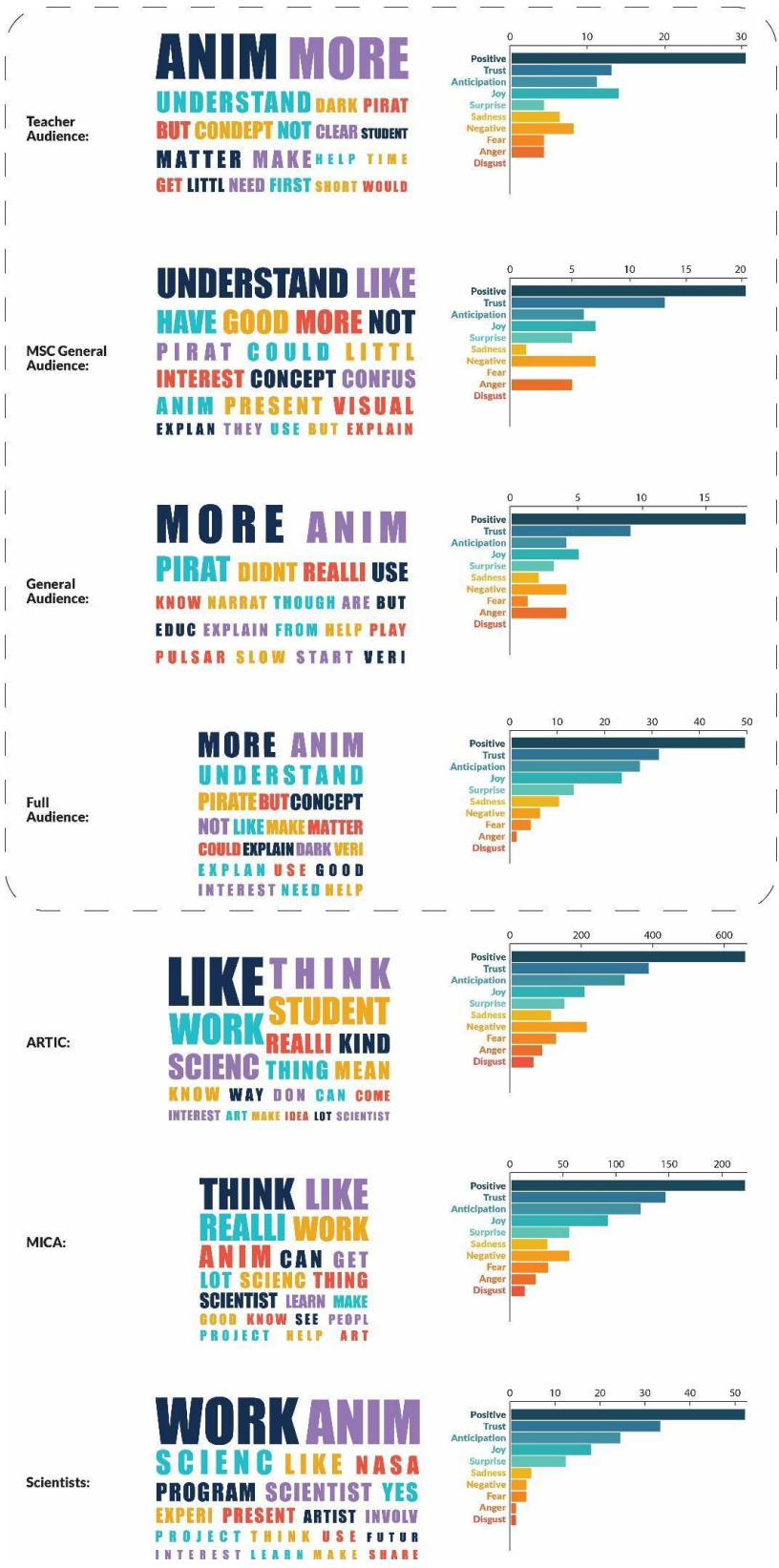

Fig. 6. Word frequency analysis. (© Laurence Arcadias)

*Students gained a lot from the class. (N = 10)*
"This experience has been very exciting and different, and taught me so much about incorporating science and astronomy into my animations."

*Mentor scientists also responded very positively. (N = 4)*
"Fantastic program. Students consistently deliver fascinating results. Communicating science to the general public is vital for both society and for scientists and this is one of the best ways to do so."

*Non-mentor scientists had valuable comments. (N = 27)*
"How effective the science part was (in terms of relaying scientific facts) varied from presentation to presentation of course. Sometimes this did not seem to be (and didn't need to be?) the main focus but rather showing that these topics can be exciting, fun, relatable."
"One of the things I really like and is fairly unique is that most of the animations concentrate on artistic presentation and response to the science and not on being a science explainer or teaching tool. This is a really valuable and appealing approach to science appreciation." However, one scientist expressed concern about the potential for science to be misunderstood.

*The school teachers were more reserved. (N = 37)*
"It needs a little bit more detail. The animations were good, but many concepts are still vague."
**"The pirate ship looks cool. Quasars + Pulsars well done + beautiful. The concepts were probably a little too esoteric to really be illustrated in the short format for many of the concepts. An initial understanding of the factors at play was needed. The creative representation of the content was entertaining, humorous and artful, but concrete references to observable phenomenon would help understanding."**
**"Animations are a bit unclear to me, the topics were not clearly stated at all. Need more detailed descriptions."**

*The general audience at the MSC expressed mixed feelings. (N = 25)*
"These are cute and definitely spark interest! I still don't understand the topic though so either people would need background knowledge to fully understand the videos or they could be used to spark interest to want to learn more."

*Audiences showed more enthusiasm when the animations were presented by both animators and scientists. (N = 15)*
From the presentations at the GSFC visitor center and the outreach event, comments included: "I thought it was a great example of interdisciplinary collaboration, and a great way to energize creativity & spark interest in science by the general public."

We show a word frequency analysis of the open-ended survey responses in Fig. 6. The most frequent terms included suggestions for improvements (e.g., "more", "but", "not", and "could") and positive feedback (e.g., "understand", "like", "good", "interest").

Our primary conclusion from the surveys was that presenting astronomy via this type of animation is highly attractive to a range of audiences. We found that simply screening the animations on their own has a visual impact and may raise scientific awareness. However, for science learning and education it is essential to provide some additional context, which may be as little as a few explanatory sentences. When this is available, then the overall responses are more positive.

### Highlights of interviews

### Mentor Scientist Interviews

Mentor scientists found the project to be an effective way to enhance science communication and outreach with the general public by offering a different perspective. They found that explaining the science concept to the students helped them to be more concise. Seeing the animation process and the range of techniques was revealing for them. Most wished to work with artists again and expressed their intention to use the animations for outreach and scientific presentations.

### Student Interviews

Some students expressed that making an animation was a very effective way to learn about a science concept. While our goal for the class was for the students to have artistic freedom to pursue less-literal approaches, a number of students felt that their animations had to be strongly constrained to "stick to the science facts". Recently, more animations have moved toward a freer interpretation.

Associating creativity with the research was inspiring to students, making them aware of their roles as artists connecting to science. Some students noted that scientists need artists to show their research in a more playful and effective way and some saw an opportunity for further art and science collaboration.

Working with astrophysicists helped the students realize that scientists have broader backgrounds than their initial ideas. Collaborating with the scientists was reported to feel relatively easy. Some students mentioned their fear or embarrassment to either not understand the scientists, or to disturb them in their work at the beginning of the project. However, these feelings changed over time and students appreciated the accessibility and the excitement of the scientists when looking at their work. One student regretted not having more time to work more closely with a scientist during the process. For most students, the experience reinforced their understanding of the scope of animation, having the potential to tell different types of stories. Some students liked the fact that this class teaches them to apply animation to other fields. The amount of work seemed challenging for a few students, but most students felt that the overall experience was very positive for them.

# Similarities and Differences Between SAIC And MICA Programs

The SAIC offers a combined studio/science course that shares some similarities with the MICA astro-animation class. For example, both programs are guided by an astrophysicist and an artist and count for six credits. However, there are some clear differences. The SAIC faculty expressed that their program's objective is situated on the art side of the art-science intersection, exhorting students to examine what happens when their artwork comes into contact with physics. In contrast, the aim of the MICA class is to push art students to greater science awareness as a source of inspiration and collaboration.

The SAIC syllabus mentions exploring research-based art practice as well as embracing the possibility of art itself as a form of research. This may give the SAIC students a certain freedom to express their ideas with a primary focus on art. The SAIC program is open to any art form while the MICA class produces exclusively animation.

The SAIC artwork has been utilized to a significantly lesser extent for science outreach than we do with our MICA class. However, the SAIC work has been successfully exhibited at the Sullivan Galleries in Chicago.

Overall, it appears that both programs have strong value for the students involved, even though one is focused on art practice and research, and the other centered on art/science collaboration.

# Discussion

## Highlights of Results

Our investigation showed that there was a very positive reception to the animations produced in the astro-animation class from both specialist and non-specialist audiences. Scientists were generally very supportive, although a few expressed concerns that sometimes the science was lost. Some audience participants commented on a lack of clarity in the science message, however, the goal for these animations is not necessarily to explain science but to trigger an interest in the subject and make it more approachable. **To communicate the underlying science more clearly, the animations would preferably be supported by additional material or presented by people with science knowledge.** This program offers an effective way for art students, not only to learn science but to have a glimpse of "science in action" and potentially become involved in the process as artists. This already has happened in the short term with several students becoming summer interns at NASA.

## Challenges

Undergraduate art education in the United States typically requires that students take a science class. By combining astronomy with animation in a single class, our course seems to be the perfect fit. Despite the popularity of such programs and the desire of our institution to move toward an integrated model it remains difficult to schedule such classes in academic settings. A project such as the astro-animation class requires a high level of passion and commitment in order to face the many challenges that may arise in bringing together such a collaboration [13].

We were initially concerned that after a few years the scientists might lose interest in our project. However, while we started our class with a small number of NASA scientists, it has been snowballing to other groups, which helped us develop new topics such as commemorating the 50th anniversary of the moon landings in 2019.

During the Covid confinement, we were still able to deliver our class online, although not being able to visit NASA physically may explain a small drop in the course registration. Student public presentations through video conferencing were very popular the first year, perhaps because of the novelty of the situation, but somewhat less well-attended in the second year, perhaps because of virtual fatigue.

## <2> Lessons Learned and Recommendations

We present here our major recommendations, see also [12] for a summary. They depend on the objective of a program. Their relative importance can differ depending on whether a program has art and science as equal components, or science is being used as a "point of entry" for art research.

**Learning goals and measurement** must be defined clearly. In our case, the science learning part was achieved by lectures, hands-on experiments, research reading and discussion with scientists. The measurement happened through weekly quizzes and a final test. The animation learning was achieved through the artistic science response and production of the animation and the measurement of the animation happens through a series of rubrics: creative voice, pertinence of response, production quality, collaboration efficiency.

**Proximity to dedicated scientists** was crucial for the class success. NASA being about 30 miles away from MICA was important to initiate our collaboration.

**Working communication between students and scientists** must be established as soon as the first day of class. In addition to the face-to-face meetings, we use Google Drive to store the work and Tumblr to show the work in progress. For direct communication, both parties use email for questions and feedback.

**Sharing the "how it's made" component of the students' animation work** with audiences and the scientists is a great way to engage with people. It brings to light the art component in addition to the science.

**Making the science class hands on** with in-class physical demonstrations helped improve the students' engagement and attention. This is a great way of understanding the physics concepts as well as encouraging students to conduct their own experiments. Having a playful attitude toward science has a direct connection with artistic exploration.

**NASA online educational resources,** in particular the Afterschool Universe [15], were found to be very valuable to augment the classroom learning experience. NASA's extensive open-source material such as the free sound images and video library offers great opportunities for the students to explore.

**Community Engagement.** The outputs of such a class have a broad popular appeal. For example, the students presented their animations at events such as the Balticon and Escape Velocity science fiction conferences. This shows the students the power of using art to communicate important messages for public awareness.

# Conclusion

The results of the surveys and interviews on the astro-animation project show very positive perceptions and a lot of enthusiasm from the scientists, the animation students, and the general public. The combination of animation together with astronomy produces very rich outcomes. Both the process of creating the animations and the resulting films have a wide variety of benefits. They facilitate the science education of the student artists, produce animations that are powerful in themselves, and encourage scientists to become more involved in communication with the general public. Moreover, the increasing number of animations available is a resource that can continue to be used for outreach. The program has a potential to grow in different directions such as STEM learning or by expanding the artistic outcomes, supporting artistic visions, and raising worldview questions. With a stronger understanding of science, students are empowered to become activists in the cause of science- and evidence-based policies.

## Acknowledgements

We thank the many student animators at MICA and scientists at NASA GSFC, JHU APL, and elsewhere who created the animations that are at the core of this program. We also received support and help from many others at MICA and NASA GSFC. We thank Jay Friedlander for extensive participation in the project. We thank Rong Chen for statistical analyses, KC Corbett for graphics production, Declan McKenna and Vy Phan for transcribing interviews, Brigitte Pocta for editing assistance, and Kathleen Ward for assistance with administering questionnaires and conducting interviews, and MICA's Office of Research. We thank the Fermi team for their support of this project since its earliest stages, and the Fermi Cake Committee for their sweet contributions. Paola Cabal, Kathryn Schaffer and their class are thanked for their hospitality at SAIC. This research was supported by a grant from the National Endowment for the Arts, number 17-980145 and NASA (80GSFC17M0002).

## Biographical Information

Laurence Arcadias is a professor in the Animation Department at the Maryland Institute College of Art. She received her MFA from the Institute of Visual Arts, Orléans, France.

Robin Corbet is a Senior Research Scientist in the Center for Space Sciences and Technology at the University of Maryland, Baltimore County, working at the NASA Goddard Space Flight Center. He received a PhD in high-energy astrophysics from University College London.